\documentclass{article}
\usepackage[utf8]{inputenc}
\usepackage{authblk}
\usepackage{graphicx}
\usepackage{cite}

\usepackage[a4paper,top=3cm,bottom=2cm,left=3cm,right=3cm,marginparwidth=1.75cm]{geometry}
\usepackage{amsmath}
\usepackage{hyperref}
\usepackage[pagewise]{lineno}

\usepackage[colorinlistoftodos]{todonotes}

\title{Estrategias de movilidad basadas en la teoría de percolación para evitar la diseminación de enfermedades: COVID-19\\
\vspace{1cm} 
Mobility strategies based on percolation theory to avoid the spread of diseases: COVID-19}
\author[1]{Diana Rosales Herrera}
\author[,2]{J. E. Ramírez\footnote{Corresponding author: jhony.ramirezcancino@viep.com.mx}}
\author[,1]{Jorge Velázquez-Castro \footnote{Corresponding author: jorgevc@fcfm.buap.mx }}
\author[3,4]{Bogar Díaz}
\author[1]{M. I. Martínez}
\author[5]{P. Vázquez Juárez}
\author[1]{A. Fernández Téllez}

\affil[1]{Facultad de Ciencias F\'isico Matem\'aticas, Benem\'erita Universidad Aut\'onoma de Puebla, Apartado Postal 165, 72000 Puebla, Pue., M\'exico}
\affil[2]{Centro de Agroecología,
Instituto de Ciencias,
Benemérita Universidad Autónoma de Puebla, Apartado Postal 165, 72000 Puebla, Pue., M\'exico}
\affil[3]{Departamento de Matem\'aticas, Universidad Carlos III de Madrid. Avda.\  de la Universidad 30, 28911 Legan\'es, España}
\affil[4]{Grupo de Teorías de Campos y Física Estadística. Instituto Gregorio Millán (UC3M), Unidad Asociada al Instituto de Estructura de la Materia, CSIC, Serrano 123, 28006 Madrid, España}
\affil[5]{Facultad de Contaduría, Benem\'erita Universidad Aut\'onoma de Puebla, Apartado Postal 165, 72000 Puebla, Pue., M\'exico}

\date{}

\begin{document}

\maketitle

\begin{abstract}
La movilidad de las personas es uno de los principales factores que propician la propagación de epidemias. En particular, es el factor que genera el esparcimiento de la enfermedad en diferentes regiones.
Las medidas de control epidemiológico basadas en la restricción de movilidad son generalmente poco populares y las consecuencias económicas pueden llegar a ser muy grandes. Debido a los altos costos de estas medidas, es de gran relevancia tener estrategias globales que optimicen las medidas minimizando los costos. 
En este trabajo, se calcula el umbral de percolación de la propagación de enfermedades en redes. De manera particular, se encuentra el número de caminos a restringir y localidades que tienen que ser aisladas para limitar la propagación global de COVID-19 en el Estado de Puebla, México.  
Simulaciones computacionales donde se implementan las medidas de restricción de movilidad entre los diferentes municipios, junto con las medidas de confinamiento, muestran que es posible reducir un 94\% de la población afectada comparado con el caso en el que no se implementa ninguna medida. Esta metodología puede ser aplicada a distintas zonas para ayudar a las autoridades de salud en la toma de decisiones. \\
\end{abstract}

\begin{abstract}
    Human mobility is an important factor in the propagation of infectious diseases. In particular, the spatial spread of a disease is a consequence of human mobility. On the other hand, the control strategies based on mobility restrictions are generally unpopular and costly. These high social and economic costs make it very important to design global protocols where the cost is minimized and effects maximized. In this work, we calculate the percolation threshold of the spread in a network of a disease. In particular, we found the number of roads to close and regions to isolate in the Puebla State, Mexico, to avoid the global spread of COVID-19. Computational simulations taking into account the proposed strategy show a potential reduction of 94\% of infections. This methodology can be used in broader and different areas to help in the design of health policies.  
    

\end{abstract}

    \textit{Descriptores:} Teoría de percolación, umbral de percolación, propagación de enfermedades , COVID-19\\


\section{Introducción}
A finales del mes de diciembre de 2019 se supo de la aparición de un nuevo virus, SARS-CoV-2, de la familia  corona, que inició su propagación en un mercado de animales de Wuhan, Hubei, China Central. Este virus provoca una enfermedad respiratoria, denominada COVID-19, altamente contagiosa y de alta letalidad en individuos con comorbilidades, de edad avanzada o con problemas respiratorios \cite{kronbichler2020asymptomatic, factoresriesgocovid}. La enfermedad se propagó en aquella región del país asiático y en pocos meses se registró la presencia de este virus en todo nuestro planeta,  convirtiéndose en una pandemia que ha causado la muerte de cientos de miles de seres humanos en todos los continentes. 
Por lo tanto, es necesario estudiar la situación actual y profundizar en posibles estrategias que pueden ayudar a mitigar la propagación espacial de este virus y otras enfermedades.
Para esto, las autoridades de diferentes países han implementado medidas locales de confinamiento, en donde se limita hasta cierto grado la movilidad de las personas. Sin embargo, éstas medidas 
han sido impuestas de manera general y al parecer sin tomar en cuenta el tipo de conexión que hay entre los poblados. Los resultados de estas medidas han sido variados y también los analizaremos para algunos países.

Por otro lado, en física estadística, la teoría de percolación ha sido ampliamente utilizada para estudiar los proceso de flujo a través de un medio poroso \cite{SABERI, souza}. 
En esta teoría, el medio poroso puede modelarse como una red cuadrada, en donde cada celda es asignada como ocupada o vacía con probabilidad de ocupación $p$ y $q=1-p$, respectivamente. 
Esta asignación se realiza de manera independiente del estado de ocupación de las celdas vecinas.
De esta forma, las celdas ocupadas resultan de interés puesto que a través de estas es que el proceso de transporte puede ocurrir.
Para valores pequeños de $p$, hay pocos sitios ocupados, por lo que el proceso de transporte no puede ocurrir.
Por otro lado, cuando $p$ toma valores cercanos a 1, en el sistema habrá muchos sitios ocupados, la mayoría de ellos agrupados en un racimo lo suficientemente grande como para conectar dos extremos del sistema. Este racimo se conoce como racimo percolante y su existencia garantiza que el proceso de transporte ocurra.
En teoría de percolación, el problema fundamental que se debe resolver es la determinación del valor mínimo de la probabilidad de ocupación al cual emerge el racimo percolante. Este valor crítico es conocido como umbral de percolación \cite{souza, stauffer}.
El rango de aplicaciones de la teoría de percolación va desde el entendimiento de las propiedades del plasma de quarks y gluones \cite{braun, string, nos} hasta la formación de galaxias \cite{Schulman}.
En particular, analizando las propiedades de conexión de los grafos formados por las interacciones sociales y las propiedades de ciertas enfermedades, es posible determinar ciertos umbrales de percolación que evitan que en el sistema emerja una epidemia que afecte a una buena parte de los individuos \cite{sp, intro3, intro4, intro2, sp4,sp2,sp3}.
Recientemente, este tipo de modelos se ha aplicado en agronomía para modelar y proponer estrategias que eviten la diseminación de fitopatógenos en plantaciones \cite{JE, JE2}.


\begin{figure}[h]
\centering
\includegraphics[scale=0.15]{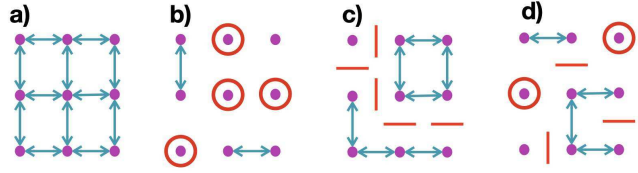}
\caption{Esquemas de los enfoques de contención de la propagación espacial de COVID-19 basados en la teoría de percolación aplicados a nueve municipios (nodos morados) enlazados (flechas azules). En a), se ejemplifica la situación de no establecer restricciones, por lo que todos los nodos se encuentran conectados a sus vecinos cercanos, propiciando la propagación espacial de la enfermedad. En b), al aislar nodos (círculos rojos) se limita la movilidad sobre dichos municipios, reduciendo así la interacción con los nodos colindantes. En c), se corta el flujo migratorio entre dos municipios conectados (líneas rojas). En d), se bosqueja una combinación de las restricciones mostradas en b) y c).}
\label{fig:prueba}
\end{figure}
En este trabajo, analizaremos las propiedades de conexión y percolación del Estado de Puebla en su representación en forma de red compleja, en donde cada sitio en la red representará un municipio, mientras que las aristas representarán la conexión entre ellos debido a la movilidad de personas.
En este sentido, acoplaremos un modelo de compartimentos para simular el proceso de propagación de COVID-19 sobre los diferentes municipios del estado.
Con base en la teoría de percolación, exploraremos algunas estrategias de confinamiento no local, como la restricción de movilidad entre municipios colindantes o sobre ciertas carreteras, que prevengan la propagación espacial de epidemias sobre la mayoría de los municipios.
De esta forma, analizamos los siguientes enfoques: i) removiendo nodos, ii) removiendo aristas, y iii) una combinación de las dos anteriores. En Fig.~\ref{fig:prueba} bosquejamos estas estrategias en una red cuadrada en dos dimensiones.
La intención de este artículo es analizar el impacto de adoptar diferentes estrategias con el fin de mitigar o evitar la diseminación de la enfermedad COVID-19. Por esta razón, no nos enfocaremos en tratar de reproducir los resultados reportados por las entidades oficiales.

El resto de este artículo está organizado de la siguiente manera. 
En Sec.~\ref{sec:percolacion} mostramos la metodología para construir la red compleja correspondiente al Estado de Puebla y sus propiedades de percolación.
En Sec.~\ref{sec:epidemiologia} evaluamos las estrategias conjuntas de confinamiento local y de restricción de movilidad basadas en la teoría de percolación con un modelo epidemiológico de compartimentos.
Finalmente, en Sec.~\ref{sec:conclusiones} discutimos los resultados más importantes y las limitaciones de las estrategias propuestas en este artículo.

\section{Propiedades de percolación del Estado de Puebla}
\label{sec:percolacion}

\subsection{Construcción de la red}

\begin{figure}
\centering
\includegraphics[scale=1]{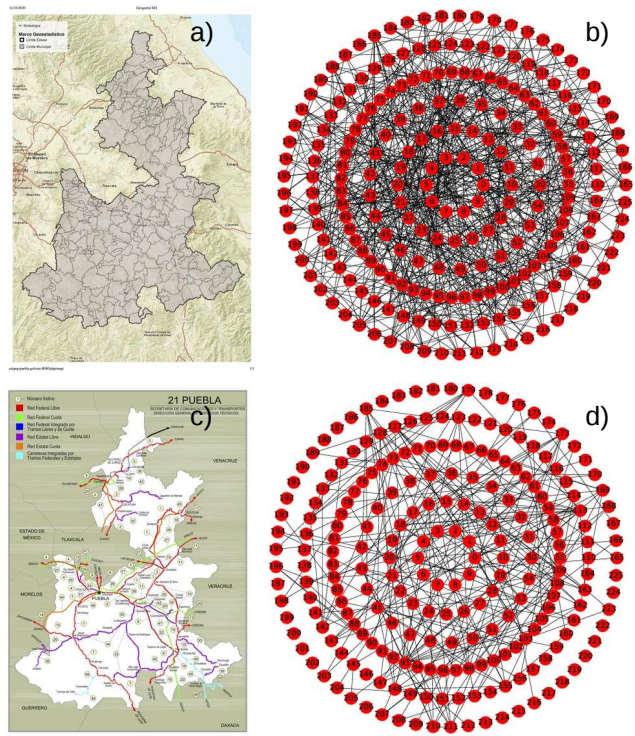}
\caption{a) Mapa de la división política del Estado de Puebla. b) Esquema en representación de la red compleja del Estado de Puebla considerando como nodos a los municipios y las aristas a las colindancias inmediatas entre los vecinos.
c) Mapa de la principal infraestructura carretera del Estado de Puebla.
d) Representación en forma de de red compleja de la infraestructura carretera del Estado de Puebla. Nuevamente los nodos representan a los municipios mientras que las aristas representan la existencia de una carretera que une a dos municipios. Se incluyen  nodos extras debido a que existen 9 municipios que geográficamente se encuentran divididos.}
\label{fig:mapa-grafo}
\end{figure}

Para estudiar la estructura de la red formada por los municipios (nodos) del Estado de Puebla y la interacción entre ellos como un sistema percolante, consideraremos dos tipos de conexión entre los municipios, siendo la  primera la interacción entre municipios colindantes. 
Para esto a cada municipio se le asigna una etiqueta según la numeración de la división político-administrativa presentada por el gobierno del Estado de Puebla (ver Fig.~\ref{fig:mapa-grafo} a)). 
Para cada uno de los 217 municipios se registraron las etiquetas de los municipios que lo rodean, teniendo especial cuidado en aquellos municipios que se encuentran divididos en dos parches, como es el caso de Acatlán de Osorio, Caltepec, Chiautla, Huehuetla, Huehuetlán el chico, Jonotla, Rafael Lara Grajales, Tecomatlán y Zoquiapan. 
Bajo estas consideraciones, el total de sitios que conforman la red del Estado de Puebla suman 226. 
Esta información puede consultarse en \cite{microrregionespueblagob}. 
La red correspondiente a este caso se presenta en la Figura \eqref{fig:mapa-grafo} b).


Para el segundo caso, se definió la conexión entre municipios considerando si existen de por medio carreteras y caminos que los conectan.
Solo conectaremos aquellos municipios por donde pase una carretera federal y estatal, tanto libre como de cuota (ver Fig.~\ref{fig:mapa-grafo} c)). 
Para el caso de las carreteras de cuota, se identificaron las etiquetas de los municipios por los cuales cruza cada tramo carretero descrito en el Informe de Datos Viales 2013 de la Secretaría de Comunicaciones y Transportes \cite{datosvialesSCT}. 
Posteriormente, todas las etiquetas de cada tramo se enlazaron entre ellas, por ejemplo, el tramo \textit{Amozoc-Perote} que cruza por Amozoc, Tepatlaxco de Hidalgo, Acajete, Nopalucan, San José Chiapa, Libres y Tepeyahualco (con etiquetas 15, 163, 1, 104, 128, 94 y 170, respectivamente), mantiene conectado cada municipio con cada uno de los restantes.
Por otra parte, para el caso de las carreteras libres, se considera la conexión entre un municipio y sus vecinos cercanos conectados a éste por medio dichas vías. 
La definición de conexión en ambos casos (carretera de cuota o libre) se elige suponiendo que la población prefiere transitar por carreteras de cuota para moverse grandes distancias, mientras que para moverse a los municipios vecinos prefiere los caminos libres de cuota.
En Fig.~\ref{fig:mapa-grafo} d) mostramos la red correspondiente para este caso.
Adicional a la información reportada por la SCT, para la simulación de éstas redes se utilizó el material presentado en \cite{anuario13} y \cite{atlasPuebla}.

\begin{figure}
\centering
\includegraphics[scale=1]{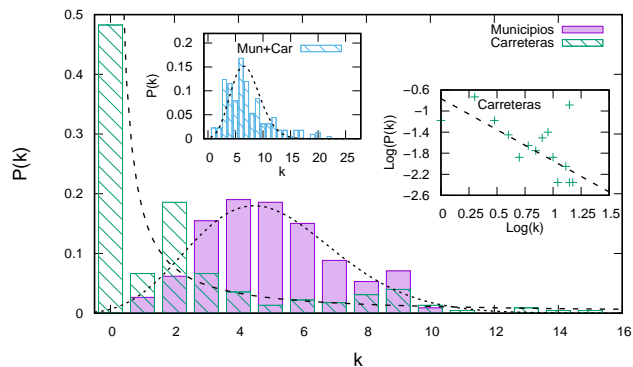}
\caption{Distribución del grado de los nodos de las redes correspondientes a los mapas de la división política del Estado de Puebla (barras moradas) y del mapa de las principales carreteras del estado (barras verdes) junto con sus respectivas tendencias (líneas negras discontinua y punteada). En el recuadro se muestra la tendencia de $P(k)$ para la red correspondiente de la infraestructura carretera en escala logarítmica.}
\label{fig:k-dist}
\end{figure}

Para clasificar de manera adecuada las redes descritas previamente, es necesario analizar la distribución del grado de conectividad de los nodos. En este caso, se debe examinar y contabilizar el número de vecinos que se encuentran conectados a cada municipio. Recordemos que dichas conexiones pueden darse por colindancias o a través de la infraestructura carretera (ver Fig.~\ref{fig:mapa-grafo}).  

En la Fig.~\ref{fig:k-dist} se muestra la función de masa de probabilidad $P(k)$ de que un municipio se encuentre conectado con $k$ nodos.
Encontramos que la red definida a través de la conectividad de las colindancias corresponde a una red tipo Erdos-Renyi \cite{newman2003structure}, con distribución de Poisson con media $\lambda\sim 5$.
Mientras que la red referente a la conexión vía infraestructura carretera corresponde a una red libre de escala, cuya distribución de conectividad es de la forma $P(k)\propto k^{-\gamma}$ con $\gamma \sim 1.2$.
De los resultados generales del umbral de percolación en redes aleatorias, para este tipo de redes es posible evitar la formación de un racimo gigante tras remover una fracción de nodos o enlaces. Por lo tanto, es factible establecer una estrategia para evitar la propagación espacial de la enfermedad basada en la teoría de percolación.
Sin embargo, para cada tipo de red debe tenerse en cuenta las implicaciones particulares de su estructura.
En la Sec.~\ref{sec:threshold} discutiremos las propiedades de percolación de la red definida a través de las colindancias y la conjunta con la red carretera.

\subsection{Umbrales de percolación}
\label{sec:threshold}

En el contexto de la teoría de percolación, una de las cantidades fundamentales que debemos analizar son el umbral de percolación, el tamaño promedio de los racimos y el tamaño del racimo gigante.
Éstas cantidades tienen una interpretación física y son de interés en el entendimiento del desarrollo del proceso de propagación de la enfermedad.
Por ejemplo, el umbral de percolación indica la fracción de nodos que es necesario remover para evitar la formación de un racimo que se extiende sobre una porción considerable del sistema. En el contexto epidemiológico de este trabajo, esto último significa que debe aislarse cierta cantidad de municipios para evitar la propagación espacial de la enfermedad sobre todo el estado. El tamaño promedio de los racimos y el tamaño del racimo gigante deben entenderse, como el número promedio de nodos afectados y el tamaño del brote de mayor extensión, respectivamente, al final del proceso de propagación.

\begin{figure}
\centering
\includegraphics[scale=1]{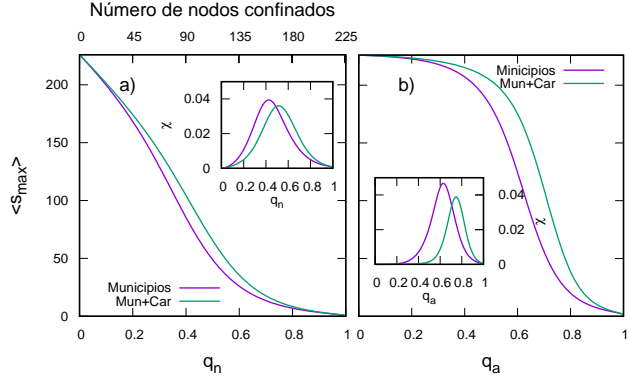}
\caption{Tamaño promedio del racimo más grande en los modelos de percolación de a) sitios y b) enlaces, considerando dos situaciones diferentes de vecinos cercanos: conexión a través de las colindancias entre municipios y una más en donde además se toma en cuenta la infraestructura carretera. En los recuadros mostramos la forma de la susceptibilidad de la red en función de los sitios/enlaces removidos.}
\label{fig:smax}
\end{figure}

Uno de los métodos ampliamente utilizados para estimar las propiedades de percolación en redes aleatorias es la simulación a través del método de Monte Carlo.
En particular, utilizamos el algoritmo descrito por Newman y Ziff \cite{newman-ziff-algoritmo}, añadiendo en la red una por una, y de manera aleatoria, las etiquetas asociadas a cada municipio. Para rastrear la formación de racimos, se verifica el estado de ocupación de todos los vecinos cercanos del nodo añadido y se {\em amalgaman} con ayuda del algoritmo \textit{Union-Find}.
Por cada sitio agregado, se mide el tamaño del racimo más grande, denotado por $s_{max}$, el cual depende de la densidad $p_n$ de nodos existentes en la red (o equivalentemente, de la fracción de nodos removidos $q_n=1-p_n$).
Debido a que los nodos tienen la misma probabilidad de ser añadidos en el $k-$ésimo paso de la simulación, es posible suavizar la curva $s_{max}$ al hacer una convolución con la distribución binomial $\mathcal{B}$, de la siguiente manera \cite{newman-ziff-algoritmo}:
\begin{equation}
    s_{max}(p_n)=\sum_{k=0}^{N_n} s_{max}(k) \mathcal{B}(N_n,k, p_n)\,,
\label{eq:conv}
\end{equation}
donde $s_{max}(k)$ es el tamaño promedio del racimo más grande tras haber añadido exactamente $k$-sitios y $N_n$ el número total de nodos en la red, en este caso 226.
En la Fig.~\ref{fig:smax} mostramos el comportamiento de $s_{max}$ como función de $q_n=1-p_n$ sobre el promedio de $10^5$ simulaciones, para los grafos construidos por colindancias (a)) y por conexión a través del sistema carretero (b)). El modelo previamente descrito es conocido como percolación de sitios, cuyo principal objetivo es estimar las propiedades percolativas del sistema tras añadir o remover cierta fracción de nodos.
En este contexto, remover un nodo en el sistema significa impedir el desplazamiento de personas hacia dentro o fuera del municipio.
La relevancia epidemiológica de éste enfoque consiste en la posibilidad de evitar la propagación de COVID-19 sobre determinada fracción de municipios al {\em aislar} de manera aleatoria cierta porción de los nodos. 
Para ser precisos, al aislar de manera aleatoria del orden del 34\% de los nodos (76 municipios), la extensión promedio de municipios afectados es del 50\% y 57\% al considerar conexión por colindancia y colindancias más sistema carretero,respectivamente.

Otra propuesta que puede emplearse es restringir el movimiento de personas sobre las aristas, es decir, evitar el flujo entre ciertos municipios sobre las carreteras que los unen.
En Fig.~\ref{fig:smax} b) se muestra el comportamiento del tamaño promedio del racimo más grande en función de la fracción de nodos removidos $q_a=1-p_a$. Para esta determinación, se añade en la red arista por arista (en lugar de nodos), y se intercambia en la Ec.~\ref{eq:conv} $N_n$ por $N_a$ igual al número total de aristas (575 y 662 para el grafo de colindancias y colindancias más carreteras, respectivamente) y $p_n$ por $p_a$. En este caso, $s_{max}(k)$ representa el tamaño promedio del racimo más grande tras haber añadido exactamente $k$ aristas.
Debido a que los grafos que representan el Estado de Puebla son altamente conectados, es necesario remover una cantidad considerable de aristas, 353 (61\%) y 460 (70\%) para el primer y segundo caso respectivamente, para reducir el promedio de municipios afectados al 50\% del total.

Adicionalmente, un enfoque más general es considerar una combinación de aislar municipios y restringir la movilidad sobre ciertos enlaces. De esta manera, se busca la combinación óptima de valores de $p_n$ y $p_a$ tales que se evite la formación de un racimo gigante de municipios afectados.
Este modelo es conocido en la literatura como percolación conjunta de nodos y enlaces \cite{hoshen2}, y los casos expuestos previamente son casos particulares de este modelo, pues en la percolación de sitios basta considerar $p_a=1$, mientras que para reproducir el modelo de percolación de enlaces se debe tomar $p_n=1$. 

\begin{table}[!ht]
\begin{center}
\begin{tabular}{c c c}
\hline
Red & $q_{nc}$ & $q_{ac}$\\
\hline
Municipios & 0.425 & 0.633\\
Municipios+carreteras & 0.513 & 0.748 \\
\hline
\end{tabular} 
\end{center}
\caption{Fracciones críticas de nodos y aristas que se requieren remover para los grafos que representan a los mapas de la división política y la infraestructura carretera del Estado de Puebla. Además se considera el grafo conjunto que se formaría entre los municipios y las carreteras.}\label{tab:treshold}
\end{table}

Por otra parte, el {\em strength} del racimo más grande puede estimarse a partir de $s_{max}$ como \cite{Radicchi}:
\begin{equation}
    P^*(p)=s_{max}(p)/N_n,
\end{equation}
mientras que la susceptibilidad de la red se calcula como \cite{Radicchi}
\begin{equation}
    \chi (p)=\frac{s_{max}^2(p)/N_n^2-[P^*(p)]^2}{P^*(p)}\,,
    \label{eq:susc}
\end{equation}
donde $s_{max}^2(p)$ debe entenderse como el promedio cuadrático del racimo más grande, y $p$ puede ser $p_n$ o $p_a$.
Una forma de estimar el umbral de percolación $p_c$ es determinando el valor de $p$ donde la susceptibilidad alcanza su valor máximo \cite{Radicchi}, es decir
\begin{equation}
    p_c=\text{arg}\{ \text{max}_p \chi(p)  \}.
\end{equation}
Los recuadros dentro de la Figs.~\ref{fig:smax} a) y b), muestran los comportamientos de la susceptibilidad en función de $q_n$ y $q_a$ para ambos grafos del Estado de Puebla. En la Tabla \ref{tab:treshold} se indica el valor de los umbrales de percolación para las redes de interés.
Debido a que la red conjunta de colindancias y carreteras está más conectada que la red de colindancias, es necesario remover más nodos o enlaces.

\begin{figure}
\centering
\includegraphics[scale=1]{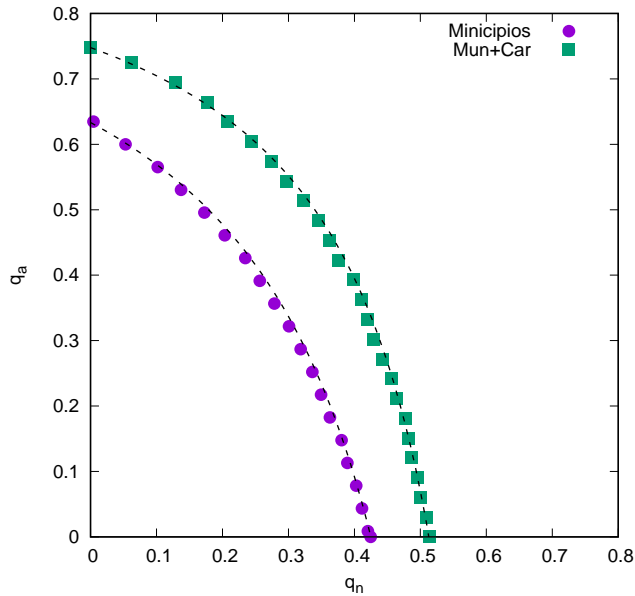}
\caption{Fracciones críticas de nodos y aristas para el modelo de percolación conjunta de nodos y aristas. Las figuras representan los resultados obtenidos a partir de la simulación por computadora, mientras que las líneas discontinuas corresponden a la parametrización dada por Ec.~\eqref{eq:para-sb} con $q_{nc}$ y $q_{ac}$ tomados de la Tabla~\ref{tab:treshold}.}
\label{fig:sb}
\end{figure}

En la Fig.~\ref{fig:sb} mostramos los umbrales de percolación en el modelo conjunto de sitios y enlaces obtenidos mediante simulación por computadora para las dos redes discutidas. 
En particular, estos umbrales pueden ser descritos a través de la siguiente parametrización \cite{tarasevich,JE2}
\begin{equation}
    q_a=1-\frac{q_{nc}(1-q_{ac})}{q_{nc}-q_{ac}q_n}\,,
    \label{eq:para-sb}
\end{equation}
donde los valores de $q_{nc}$ y $q_{ac}$ corresponden con los reportados en la Tabla~\ref{tab:treshold} para cada una de las redes.
La gran ventaja que tiene este enfoque es la posibilidad de combinar la remoción de nodos y enlaces para evitar el confinamiento de un gran número de municipios, o el cierre de todas las vías de comunicación. Por ejemplo, si $q_n=0.3$ ($\sim$ 68 municipios), es necesario evitar el flujo de personas en el 34\% y 55\% de las conexiones que conforman las redes de colindancias y el conjunto con la infraestructura carretera, respectivamente, y así evitar que se forme un racimo gigante de municipios con personas afectadas por la enfermedad.

\section{Impacto de las estrategias de confinamiento en la evolución epidemiológica}
\label{sec:epidemiologia}

Para evaluar el efecto de las estrategias de control que se derivan del estudio de percolación, se harán simulaciones numéricas de diversos escenarios.

Primero propondremos un modelo epidemiológico compartimental que constará de 6 grupos de individuos: los susceptibles $S$, los expuestos $E$, los asintomáticos $A$, los infecciosos $I$, los recuperados $R$ y los muertos $D$. Los individuos del grupo de susceptibles $S$ son las personas que potencialmente pueden adquirir la enfermedad. Los expuestos $E$, son las personas que han sido contagiadas, y por lo tanto pueden transmitir la enfermedad, pero que todavía no presentan síntomas. Las personas asintomáticas $A$ han sido infectadas pero no muestran indicios de la enfermedad durante todo el proceso de infección. La población infecciosa $I$ presenta los síntomas comunes de la enfermedad, y por lo tanto se asumen en confinamiento. Finalmente, los individuos recuperados $R$ han pasado ya por todo el proceso de infección y desarrollaron por lo menos inmunidad temporal. 

Tomando en cuenta estos grupos, el modelo local epidemiológico es
\begin{subequations}
\begin{align}
\frac{\mathrm{d}S}{\mathrm{d}t} &= -\beta S\frac{E}{N} - \beta S\frac{A}{N} - \beta \eta S\frac{I}{N} \,,\\
\frac{\mathrm{d}E}{\mathrm{d}t} &=  \beta S\frac{E}{N} + \beta S\frac{A}{N} +\beta \eta S\frac{I}{N}  - \sigma E \,, \\
\frac{\mathrm{d}A}{\mathrm{d}t} &= m\sigma E - \gamma A \,, \\
\frac{\mathrm{d}I}{\mathrm{d}t} &= (1-m)\sigma E - \gamma I \,,\\
\frac{\mathrm{d}R}{\mathrm{d}t} &= \gamma (A + (1-\mu)I) \,,\\
\frac{\mathrm{d}D}{\mathrm{d}t} &= \gamma \mu I \,,
\end{align}\label{eq:epi1}
\end{subequations}
donde $\beta$ es el riesgo de infección, y $\eta$ es la reducción en riesgo de infección debido al aislamiento parcial que tienen los individuos infectados que presentan síntomas. Por otro lado, $\sigma$ es la tasa de transición de expuestos a infecciosos, $m$ es la fracción de expuestos que evolucionan a asintomáticos, y $1/\gamma$ es el tiempo característico de la etapa infecciosa. $\mu$ es la fracción de enfermos que mueren por la enfermedad. El número reproductivo básico para este modelo es 
\begin{equation}
    R_{o}=\beta\left(\frac{1}{\sigma}+\frac{m+\eta(1-m)}{\gamma}\right) \,.
\end{equation}

Para el modelo metapoblacional indexamos a las poblaciones de cada región con el sub-índice $i \in \lbrace 1,...,n \rbrace$, y basados en un esquema Lagrangiano \cite{poletto_human_2013,bichara_sis_2015,Vel_zquez_Castro_2018} la movilidad de los individuos entre regiones está descrita por la matriz de tiempo de residencia $P=\left(p_{ij}\right)_{i,j=1}^{n}$ donde $p_{ij}$ representa la fracción de la población de la zona $i$ que en promedio se encuentra en cualquier instante en la zona $j$. 

Debido a la movilidad, el número promedio de individuos que se encuentran en cualquier momento en la región $i$ está dado por $w_{k}=\sum_{j=1}^{n}p_{jk}N_{j}$. De manera similar $\mathcal{E}_{k}=\sum_{j}^{n}p_{jk}E_{j}$, $\mathcal{A}_{k}=\sum_{j}^{n}p_{jk}A_{j}$ y $\mathcal{I}_{k}=\sum_{j}^{n}p_{jk}I_{j}$ corresponden al total de expuestos, asintomáticos e infectados en la zona $k$. Tomando lo anterior en consideración, el modelo metapoblacional correspondiente es

\begin{subequations}\label{eq:MetaPob}
\begin{align}
\frac{\mathrm{d}S_{i}}{\mathrm{d}t} &= -\beta \sum_{k}  S_{i}\frac{p_{ik}}{w_{k}}(\mathcal{E}_{k} + \mathcal{A}_{k}+\eta\mathcal{I}_{k}) \,, \\
\frac{\mathrm{d}E_{i}}{\mathrm{d}t} &= \beta \sum_{k}  S_{i}\frac{p_{ik}}{w_{k}}(\mathcal{E}_{k} + \mathcal{A}_{k}+\eta\mathcal{I}_{k}) - \sigma E_{i}\,,\\
\frac{\mathrm{d}A_{i}}{\mathrm{d}t} &= m\sigma E_{i} - \gamma A \,,\\
\frac{\mathrm{d}I_{i}}{\mathrm{d}t} &= (1-m)\sigma E_{i} - \gamma I_{i}\,,\\
\frac{\mathrm{d}R_{i}}{\mathrm{d}t} &= \gamma (A_{i}+(1-\mu)I_{i})\,,\\
\frac{\mathrm{d}D_{i}}{\mathrm{d}t} &= \gamma\mu I_{i}\,,
\end{align}\label{eq:epi2}
\end{subequations}
donde la suma sobre $k$ toma en cuenta las infecciones de los residentes de la zona $i$ en todas las posibles regiones.

\subsection{Comportamiento de $\beta$ en condiciones de confinamiento}

Para resolver las Ecs.~\eqref{eq:epi1} y \eqref{eq:epi2}, en principio se considera que los parámetros epidemiológicos 
 $\beta$, $\gamma$, $\sigma$ y $\mu$ son constantes.
Sin embargo, existen fenómenos que pueden modificar la forma en la que la epidemia evoluciona.
Por ejemplo, el análisis de los datos sobre la propagación de COVID-19 en China, Italia y España, indican que tras el confinamiento obligatorio de la población, la tasa de transmisión $\beta$ deja de ser constante para depender del tiempo $t$ de la siguiente forma \cite{caccavo2020chinese} \textcolor{cyan}{:}
\begin{align}\label{eq:betat}
\beta(t)= \begin{cases} \beta_0 & \text{sin confinamiento } \\
\beta_0 \exp (-t/\tau_\beta) & \text{con confinamiento}
\end{cases},
\end{align}
 El término $\beta_0$ corresponde a la tasa de transmisión inicial, el cual decrece exponencialmente como función de $t$ debido al efecto del confinamiento.

Para determinar el comportamiento de la tasa de transmisión $\beta$ como función del tiempo $t$ es necesario analizar los datos reportados por las instituciones.

\begin{figure}
\centering
\includegraphics[scale=1]{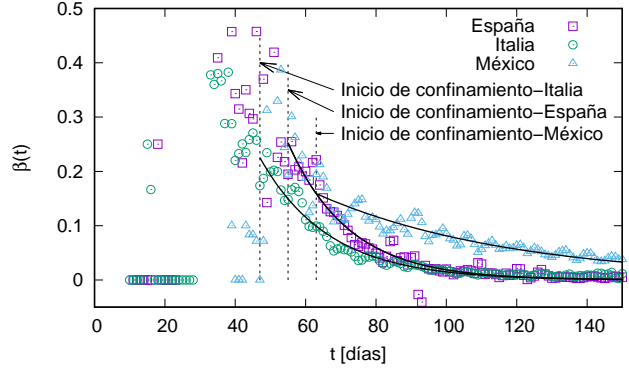}
\caption{Evolución de $\beta$ en los países España (cuadrados), Italia (círculos) y México (triángulos). El instante de tiempo $t=0$ corresponde con la fecha 22/enero/2020. Las líneas punteadas verticales corresponde con las fechas en las que se inició el confinamiento en España ($t=55$, 15/marzo/2020), Italia ($t=47$, 8/marzo/2020) y México ($t=63$, 23/marzo/2020). Las líneas negras continuas muestran las tendencias del comportamiento de $\beta$ como función de $t$ (ver Ec.~\eqref{eq:betat}) y del confinamiento establecido en cada país.}
\label{fig:beta}
\end{figure}

De manera genérica, los datos que se reportan son:
\begin{enumerate}
\item El número acumulado de casos confirmados, el cual denotaremos como $I_\text{ac}$;
\item El número de individuos recuperados $R$;
\item El número de individuos muertos $D$.
\end{enumerate}

Al no contar con la información de la transición al estado de exposición (E), usaremos un modelo SIRD para analizar los datos, el cual se expresa como\textcolor{cyan}{:}
\begin{subequations}\label{eq:SIRD}
\begin{align}
\frac{\mathrm{d} S}{\mathrm{d} t} & = -\frac{\beta}{N}SI\,, \label{eq:beta}\\
\frac{\mathrm{d} I}{\mathrm{d} t} & = \frac{\beta}{N}SI-\gamma I\,,\\
\frac{\mathrm{d} R}{\mathrm{d} t} & = \gamma (1-\mu) I\,,\\
\frac{\mathrm{d} D}{\mathrm{d} t} & = \gamma \mu I\,.
\end{align}
\end{subequations}
Es importante notar que la variable $I$ en las Ecs.~\eqref{eq:SIRD} difiere de los datos reportados $I_\text{ac}$. Sin embargo, estas cantidades se encuentran relacionadas de la siguiente manera:
\begin{equation}
    I(t)=I_\text{ac}(t)-R(t)-D(t)\,.
\label{eq:I-Iac}
\end{equation}
Mientras que la población susceptible se expresa como
\begin{equation}
    S(t)=N-I_\text{ac}(t)\,.
\end{equation}
Notamos que la expresión anterior queda completamente determinada a través de los datos observados y reportados, además la población total se conserva, $S+I+R+D=N$.

Usando las ecuaciones \eqref{eq:beta} y \eqref{eq:I-Iac}, se puede estimar el comportamiento de $\beta$ como
\begin{equation}
    \beta(t)=-\frac{N}{I_\text{ac}(t)-R(t)-D(t)}\frac{\mathrm{d} \ln [S(t)]}{\mathrm{d} t}\,.
\end{equation}
En Fig.~\ref{fig:beta}, mostramos el análisis realizado para $\beta$ usando los datos reportados para España \cite{WHOspain}, Italia \cite{WHOitaly} y México \cite{WHOmexico}. Claramente se observa una tendencia decreciente después de la fecha en donde se impone el confinamiento de la población. En este caso, $t=0$ corresponde con la fecha 22/enero/2020, a partir de la cual la Organización Mundial de la Salud lleva registro sobre los datos reportados por los organismos internacionales \cite{WHOreports}.
Como en este caso la implementación del confinamiento no coincide con el inicio de la epidemia, conviene hacer el cambio de variable $t\to(t-t_0)$ en la Ec.~\eqref{eq:betat}, siendo $t_0$ la fecha (en días contados a partir del 22/enero/2020) en la que se impuso la cuarentena. En la Tabla~\ref{tab:betafit}, se muestran los valores hallados para los datos reportados de España, Italia y México. 

\begin{table}[!ht]
\begin{center}
\begin{tabular}{c c c c c}
\hline
País &$\beta_0$ & $\tau_\beta$ & $t_0$ & Clasificación \\
\hline
España & 0.253(7) & 17.3(7) & 55 & Confinamiento estricto \\
Italia & 0.225(4) & 19.5(6) & 47 & Confinamiento estricto \\
México & 0.159(4) & 55(2) & 63 & Confinamiento relajado\\
\hline
\end{tabular} 
\end{center}
\caption{Datos de los parámetros del ajuste de $\beta$ como función de $t$ en condiciones de confinamiento en diferentes países.  \label{tab:betafit}}
\end{table}

De los resultados para $\tau_\beta$, podemos clasificar los confinamientos observados en los países analizados. Tengamos en cuenta que la interpretación de $\tau_\beta$ es tal que para el instante $t=3\tau_\beta$, $\beta_0$ disminuye su valor inicial hasta el 90\%. 
Además, el valor de $\tau_\beta$ encontrado para México es más que el doble que el encontrado para Italia y España. De esta forma, se clasifica como confinamiento estricto a las medidas implementadas en España e Italia, mientras que las que se implementaron en México como confinamiento relajado.

\subsection{Simulaciones numéricas}
\begin{figure}
    \centering
    \includegraphics{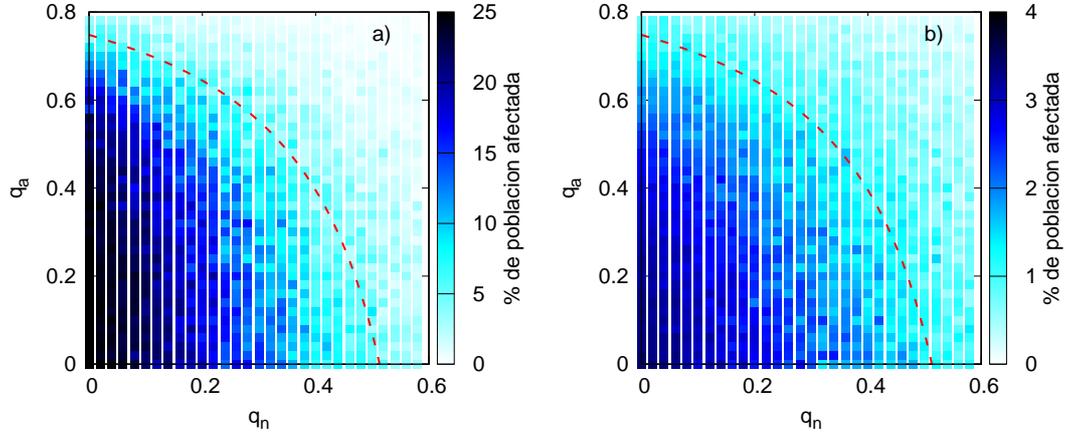}
    \caption{Porcentaje de población afectada como función de la fracción de nodos ($q_n$) y enlaces ($q_a$) removidos considerando los escenarios: (a) si no se establecen restricciones y (b) implementando medidas de confinamiento. La línea roja discontinua representa las condiciones de la curva crítica en Ec.~\eqref{eq:para-sb}.}
    \label{fig:afectados}
\end{figure}

\begin{figure}
    \centering
    \includegraphics{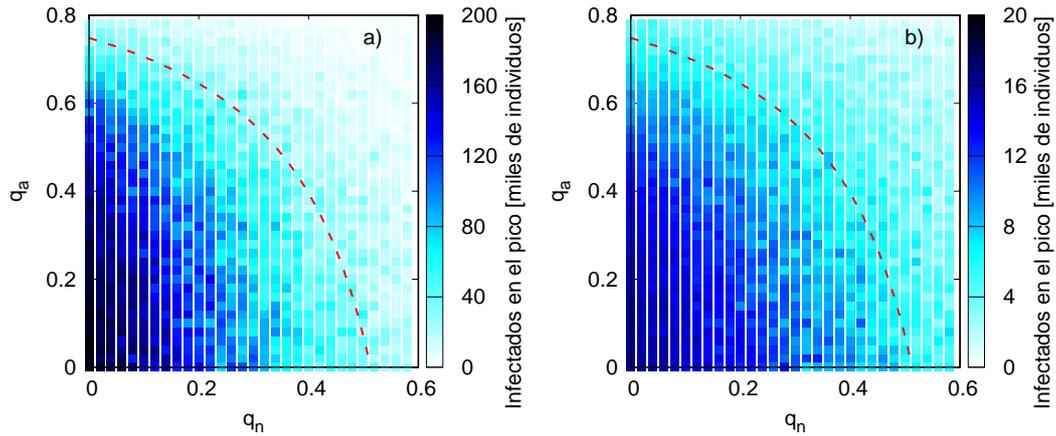}
    \caption{Número de individuos afectados en el pico de la epidemia como función de la fracción de nodos ($q_n$) y enlaces ($q_a$) removidos considerando los escenarios: (a) si no se establecen restricciones y (b) implementando medidas de confinamiento. La línea roja discontinua representa las condiciones de la curva crítica en Ec.~\eqref{eq:para-sb}.}
    \label{fig:max}
\end{figure}

Para evaluar la efectividad de las medidas de control, tanto el confinamiento local como el aislamiento de municipios y cancelación de caminos, se realizaron simulaciones numéricas empleando el modelo epidemiológico metapoblacional (\ref{eq:MetaPob}). Para lograr una comparación en distintos escenarios se fue cambiando la fracción de municipios aislados $q_{n}$ y caminos cancelados $q_{a}$. Para cada valor elegido de $q_{n}$ y $q_{a}$ se realizaron 2000 simulaciones variando aleatoriamente el lugar del inicio de la infección y los flujos de movilidad entre los caminos. La probabilidad de que la infección inicial ocurriera en determinado municipio se eligió proporcional a su población, además, se asumió que en promedio el 20\% de la población en cada localidad viaja a otros municipios. Se inició con valores $q_{n}=0$, $q_{a}=0$ y se fueron incrementando en $\Delta q_{n}=\Delta q_{a}=0.02$ hasta llegar a un valor de $q_{n}=0.6$ y $q_{a}=0.8$. De las 2000 simulaciones que se realizaron para cada uno de estos escenarios, se calculó el promedio del valor máximo de la curva de infección y el total de infectados durante toda la epidemia (ver Figs.~\ref{fig:afectados} y \ref{fig:max}). 
Además, para evaluar el efecto de las medidas de confinamiento de la población, resolveremos el modelo epidemiológico considerando el valor de $\beta$ constante (ver Figs.~\ref{fig:afectados} a) y \ref{fig:max} a)) y como función del tiempo de acuerdo con Ec.~\eqref{eq:betat} (ver Figs.~\ref{fig:afectados} b) y \ref{fig:max} b)).

\vspace{1cm}
\begin{table}
\begin{center}
\begin{tabular}{c|c|c}

Parámetro & Significado & Valor \\
\hline
 \hline
 $\sigma$  & $1/\sigma$ tiempo caractarístico en etapa expuesto & 0.265   \\
 $\mu$ & Mortalidad que genera la enfermedad  & 0.02  \\
 $m$ & Fracción de asintomáticos & 0.5  \\
 $\eta$ & Fracción de sintomáticos que propagan la enfermedad & 0.2  \\
 $\beta$ & Tasa de contacto de la población &  $\frac{\gamma Ro}{\eta(1.-m)+m+\gamma/\sigma}$  \\
 $Ro$ & Número reproductivo básico & 3.25 
\end{tabular}
\end{center}
\caption{Parámetros epidemiológicos en Ecs.~\eqref{eq:epi2} y los valores usados en las simulaciones.}  \label{tab:parametros}
\end{table}


\section{Discusión y conclusiones}
\label{sec:conclusiones}

En este trabajo exploramos los efectos que tienen diferentes medidas de confinamiento y restricciones en la evolución de la epidemia generada por COVID-19. Por ejemplo, la implementación de medidas de confinamiento de los individuos en sus hogares, o la implementación de restricciones de movilidad de las personas entre diferentes municipios colindantes o conectados a través de la principal red carreta del estado.

Como es de esperarse, la evolución de la epidemia depende fuertemente de las medidas y restricciones tomadas.  
Del análisis de los datos reportados por las autoridades, podemos concluir que en el caso de los países europeos, las medidas de confinamiento estrictas lograron disminuir la transmisión de la enfermedad en aproximadamente la tercera parte del tiempo que en el caso de México.
Cabe mencionar que la importancia del grado de confinamiento influye directamente en la duración de su implementación, y por consecuencia, en las repercusiones económicas.


La constante movilidad de las personas entre los diferentes municipios hace que la propagación espacial de las enfermedades se agudice. En este sentido, analizamos la conectividad de los municipios en el Estado de Puebla considerando las colindancias entre ellos y la infraestructura carretera que los une.
Para hacer esto, representamos al Estado de Puebla como una red de puntos (municipios) conectados a través de aristas, las cuales representan una vía de flujo migratorio entre municipios.
En Fig.~\ref{fig:mapa-grafo} mostramos la representación en forma de grafo del Estado de Puebla considerando que dos municipios están conectados si: i) son colindantes, y ii) existe una vía carretera que los vincula.
Una estrategia para evitar la propagación espacial de COVID-19 (o cualquier otra enfermedad de características semejantes) consiste en evitar el flujo de personas sobre ciertos municipios. Para esto, estudiamos desde la perspectiva de la teoría de percolación, la formación de racimos de municipios bajo las condiciones previamente mencionadas (ver Fig.~\ref{fig:mapa-grafo}). 
De esta forma, analizamos los siguientes enfoques: i) removiendo nodos, ii) removiendo aristas, y iii) una combinación de las dos anteriores. En este contexto, remover un nodo significa evitar el flujo de personas sobre un municipio particular, mientras que remover una arista equivale a evitar el flujo de personas sobre un par de municipios previamente conectados. En Fig.~\ref{fig:prueba} mostramos los esquemas de dicho enfoque considerando nodos dispuestos en una red cuadrada, donde el flujo entre municipios vecinos es señalado mediante flechas. Notar que en caso de no establecer medidas de restricción (ver Fig.~\ref{fig:prueba} a)), la propagación de la enfermedad que inicia en alguno de los nodos pasará a los vecinos cercanos, y así sucesivamente. De esta manera, la enfermedad se propagará sobre todo el grafo, o equivalentemente, sobre todo el Estado de Puebla, tal como se observa al día de hoy, en donde en todos los municipios hay reportes de personas enfermas de COVID-19.
Por otra parte, al considerar las estrategias de restricción del flujo de personas, es posible mitigar el proceso de propagación de la enfermedad, permitiendo que ésta solo se disemine sobre racimos finitos, y de esta forma, es posible proteger algunos municipios, tal como se bosqueja en Fig.~\ref{fig:prueba} b), c) y d), donde se ejemplifican las estrategias de remoción de nodos, aristas y una combinación de ambas, respectivamente.

En este estudio, las cantidades más importantes que se determinaron fueron los valores de la fracción de nodos/aristas que deben ser removidos para evitar la formación de un racimo ``gigante" de municipios afectados, los cuales fueron determinados vía simulación por computadora.
Los resultados obtenidos son los siguientes:
deben removerse el 42.5\% y 51.3\% de nodos para el grafo de colindancias y el conjunto con la estructura carretera, respectivamente. Mientras que en el caso de remover aristas, encontramos que se requiere restringir el flujo sobre el 63.3\% y 74.8\%  para el caso del grafo de colindancias y el conjunto con la estructura carretera, respectivamente.
Adicionalmente, el análisis sobre el tamaño promedio del racimo más grande en función de la fracción de nodos/aristas removidos es sumamente importante, ya que determina en promedio la extensión de los municipios afectados. 
En particular, observamos que aislar el 34\% de los municipios permite una reducción del 50\% y 43\% de municipios afectados en los casos de considerar la movilidad entre municipios colindantes y el grafo conjunto con la infraestructura carretera. Sin embargo, debido a la alta conectividad del Estado de Puebla, es necesario remover un número considerable de aristas para observar el mismo efecto.
Por otra parte, nuestro enfoque permite determinar la combinación de la fracción de nodos y aristas que se requiere remover para evitar la formación del racimo ``gigante".

En Figs.~\ref{fig:afectados} y \ref{fig:max} mostramos los resultados obtenidos para el proceso de propagación de la enfermedad considerando diferentes escenarios de medidas de confinamiento.
En Fig.~\ref{fig:afectados} a), mostramos los resultados del porcentaje de población afectada bajo el escenario de solo implementar medidas de restricción de movilidad.
Tengamos en cuenta que $q_n=0$ y $q_a=0$ representan el caso en donde no se ha impuesto ninguna medida para evitar la diseminación de la enfermedad.
En este caso, la epidemia tendrá un alcance del orden del 25\% de la población con un pico del orden de 200mil individuos afectados, lo que indudablemente saturaría la capacidad de los servicios hospitalarios en el estado.
Por otro lado, cuando se imponen medidas de restricción de movilidad entre los municipios sobre las condiciones de la curva crítica, se observa una disminución del porcentaje de personas afectadas hasta un 7\%, lo que equivale a una reducción del 72\% de la población afectada al restringir la movilidad de las personas.
Por otro lado, la implementación de medidas de confinamiento permite salvaguardar a la población de la transmisión de la enfermedad.
Un componente fundamental que contribuye a aliviar la presión en los hospitales debido a la cantidad de personas con afectaciones graves proviene de imponer a la población medidas de confinamiento, tal como mostramos en Fig.~\ref{fig:afectados} b).
Esto permite una reducción de manera eficiente en el porcentaje de personas afectadas hasta un 4\%, incluso en situaciones en donde no se establecen medidas en las restricciones de movilidad entre los municipios. 
Más aún, cuando se aplican tales medidas, el porcentaje de personas afectadas sobre la curva crítica incluso se reduce hasta un 1.5\% de la población, lo cual representa una reducción del orden del 63\% en el número de casos, y comparado con el peor escenario (no imponer ningún tipo de medidas), la reducción toma valores del orden del 94\%.

Cabe resaltar que los resultados aquí obtenidos han sido analizados considerando la remoción aleatoria de nodos y aristas de los grafos previamente descritos. Sin embargo, si se permite que los individuos se desplacen más allá de los primeros vecinos aquí definidos, lo que sucederá es que la conectividad de los grafos se verá aumentada, lo que causaría que la cantidad de nodos y aristas que se requiere remover para mantener los resultados aquí comentados sea más alta. Es por ello la relevancia de atender la sugerencia de evitar la movilidad de la población sobre los diferentes municipios. 

Uno de los problemas sociales fundamentales que atendemos en esta propuesta consiste en la limitación del flujo migratorio entre municipios, sin que se altere o se vea modificada la cadena de suministros. Para esto, es necesario mantener un control estricto sobre el estado de salud del personal encargado del transporte de alimentos, medicinas, suministros, etc. y no sobre toda la población.

Finalmente, aún quedan algunas preguntas abiertas. Por ejemplo: ¿Durante qué tiempo es necesario mantener el aislamiento? Es claro que dentro de los municipios que se encuentren en aislamiento sería posible que las personas se desplacen libremente, sin embargo, en los municipios en los que se permite el tránsito de personas con otros municipios, ¿sería conveniente imponer medidas de confinamiento? ¿Estricto o relajado? ¿Durante cuánto tiempo? 
Este modelo puede ser aplicado a cada uno de los estados del país o extendido para analizar en su totalidad los municipios que lo componen.

\section*{Agradecimientos}
Agradecemos el apoyo económico por parte del Consejo de Ciencia y Tecnología del Estado de Puebla. Bogar Díaz está financiado por el programa CONEX-Plus, con fondos de la Universidad Carlos III de Madrid, del Programa Horizonte 2020 dentro de las acciones de trabajo Marie-Sklodowska Curie COFUND (H2020-MSCA-COFUND-2017- GA 801538).

\bibliographystyle{ieeetr}
\bibliography{main}
\end{document}